\documentclass[acmsmall]{acmart}
\AtBeginDocument{%
  }

\setcopyright{cc}
\setcctype{by}
\acmJournal{PACMHCI}
\acmYear{2025} \acmVolume{9} \acmNumber{7} \acmArticle{CSCW519} \acmMonth{11} \acmPrice{}\acmDOI{10.1145/3757700}

\usepackage{xcolor}
\newcommand{\edited}[1]{\textcolor{black}{#1}}
\newcommand{\minor}[1]{\textcolor{black}{#1}}

\usepackage{enumitem}

\usepackage{tabularray}
\AtBeginDocument{%
  }

\begin{document}

\title[Opportunistic Interactions in Social VR]{Verisimilitude as Boon and Bane: How People Initiate Opportunistic Interactions at Professional Events in Social VR}

\author{Victoria Chang}
\email{vchang@umd.edu}
\affiliation{%
  \institution{College of Information,
  University of Maryland}
  \city{College Park}
  \state{MD}
  \country{USA}
}

\author{Caro Williams-Pierce}
\email{carowp@umd.edu}
\affiliation{%
  \institution{College of Information,
  University of Maryland}
  \city{College Park}
  \state{MD}
  \country{USA}
}

\author{Huaishu Peng}
\email{huaishu@umd.edu}
\affiliation{%
  \institution{Department of Computer Science,
  University of Maryland}
  \city{College Park}
  \state{MD}
  \country{USA}
}

\author{Ge Gao}
\email{gegao@umd.edu}
\affiliation{%
  \institution{College of Information,
  University of Maryland}
  \city{College Park}
  \state{MD}
  \country{USA}
}

\renewcommand{\shortauthors}{Victoria Chang, Caro Williams-Pierce, Huaishu Peng, and Ge Gao}

\begin{abstract}
Opportunistic interactions—the unstructured exchanges that emerge as individuals become aware of each other's presence—are essential for relationship building and information sharing in everyday life. Yet, fostering effective opportunistic interactions has proven challenging, especially at professional events that have increasingly transitioned from in person to online formats. In the current paper, we offer an in-depth qualitative account of how people initiate opportunistic interactions in social VR. Our participants consisted of 16 individuals with ongoing experience attending VR-mediated events in their professional communities. We conducted extensive observations with each participant during one or more events they attended. We also interviewed them after every observed event, obtaining self-reflections on their attempts to navigate opportunistic interactions with others. Our analysis revealed that participants sought to understand the extent to which social VR preserved the real-world meanings of various nonverbal cues, which we refer to as verisimilitude. We detailed the unique connections between a person's perceived verisimilitude and their social behaviors at each of the three steps toward initiating opportunistic interactions: availability recognition, attention capture, and ice-breaking. Across these steps, the VR platform typically replaces complex social mechanisms with feasible technical ones in order to function, thereby altering the preconditions necessary for a nonverbal cue's social meanings to remain intact. We identified a rich set of strategies that participants developed to assess verisimilitude and act upon it, while also confirming a lack of systematic knowledge guiding their practices. Based on these findings, we provide actionable insights for social VR platform design that can best support the initiation of opportunistic interactions for professional purposes.
\end{abstract}



\maketitle
\section{Introduction}
Social VR has the potential to expand human interaction possibilities, reaching beyond what traditional computer-mediated mediums can offer~\cite{mcveigh2019shaping, mcveigh2018s}. By combining the spatial immersion of VR technology with the social connectivity of online platforms, people can experience the lifelike sense of ``being together'' while remaining physically apart~\cite{freeman2022working}. These affordances pave the way for more natural and fluid social behaviors within virtual environments.

The current research connects social VR with \textit{opportunistic interactions in professional settings}---the unstructured yet essential exchanges that enable social bonding and knowledge transfer among members of the same workspace or community. Previous work in CSCW and communication research has delved deeply into each of these areas. Behavioral studies on how people coordinate spontaneity in workplace communication can date back to early 1990s, followed by two decades of experimental work aiming to reproduce this coordination in remote settings (e.g., ~\cite{tang1994montage, fish1990videowindow, isaacs1993video}). Under a separate endeavor, recent scholarship has explored the emerging use of social VR for social hangouts and networking in informal meetups (e.g., ~\cite{mcveigh2019shaping, moustafa2018longitudinal, maloney2020falling}). Little research, however, exists at the intersection of these two, although social VR arguably holds greater promise than most other mediums for supporting high-stakes interactions. 

We offer an in-depth qualitative account to address this oversight. Specifically, we investigate factors that contribute to the successes and failures of opportunistic interactions in VR-mediated professional events. All participants involved in this investigation were authentic users of social VR. With their consent, we shadowed each person during one or more public professional events they have signed up for, closely observing the way they navigated each opportunistic interaction in its social context. We also conducted post-event interviews with them to gather their retrospections, enabling triangulation. The data obtained from these steps allow us to develop a nuanced understanding of how interaction unfolds in a context jointly characterized by spontaneity, professionalism, and virtuality.

Our findings provide rich empirical insights that solidify the promise and current roadblocks toward effectively initiating opportunistic interaction in social VR. At the center of these findings lies the notion of verisimilitude, or the participant’s perceived extent to which a social VR environment has preserved the real-world meanings of various non-verbal cues exchanged there. Our participants’ behaviors and interview responses revealed three distinct steps leading to the eventual occurrence of opportunistic interaction in professional settings: availability recognition, attention capture, and ice-breaking. At each step, an individual's assessment of verisimilitude could be disrupted by a varied set of issues. Participants demonstrated an escalating need for high verisimilitude as they advanced through the three steps. They showed an increasing reliance on the host’s assistance when this need was not met. These insights provide actionable guidance for the design of future social VR platforms, as well as for people who navigate professional interactions via those platforms.

\section{Related Work}
This section summarizes prior studies and findings related to our current research. We begin by reviewing literature on opportunistic interactions in traditional face-to-face workplaces, emphasizing its crucial role in professional settings (Section~\ref{opportunistic interactions_F2F}). Next, we compare the similarities and differences between face-to-face and video-mediated opportunistic interactions for professional purposes (Section~\ref{opportunistic interactions_Online}). We then synthesize existing insights on social VR as an emerging medium for opportunistic interactions, primarily in casual or recreational contexts. Together, these three strands of literature set the stage for our investigation of opportunistic interactions in professional social VR settings (Section~\ref{opportunistic interactions_SVR}).

\subsection{Opportunistic Interactions in Professional Face-to-Face Settings}\label{opportunistic interactions_F2F}
Opportunistic interactions play a crucial role in professional face-to-face settings~\cite{kraut1990informal, fay2011informal}. Studies show that in office environments, opportunistic interactions can constitute up to approximately 92\% of all conversations~\cite{whittaker1994informal}, fostering strong social bonds among colleagues~\cite{holmes2004relational}, enabling spontaneous problem-solving without the formal pressures of meetings~\cite{conrad1997introduction}, and helping to build meaningful interpersonal relationships~\cite{duck1993social}. Beyond the workplace, opportunistic interactions is equally essential in professional contexts like conferences and workshops, where it facilitates networking, secures business clients and contracts, and promotes academic collaborations~\cite{oester2017conferences, mccarthy2004augmenting}.

Despite the prevalence of opportunistic interactions in professional settings, initiating these interactions can feel challenging due to the complex and dynamic behaviors involved~\cite{pillet2018begin}. Research in communication and psychology has therefore examined how face-to-face conversations are initiated, identifying two main features: audible and non-verbal cues~\cite{kendon1973greetings, pillet2012greeting}. For example, Pillet-Shore analyzed 337 video-recorded openings of face-to-face interactions, finding that people often use greeting utterances like \textit{hello }or \textit{good morning} to acknowledge another's presence as a conversation starter. While these greetings may appear similar, they carry subtle nuances along a prosodic continuum, with ``large'' greetings expressing positivity and ``small'' ones conveying a neutral stance~\cite{pillet2012greeting}.

Non-verbal signals are also essential and often function in combination with audible cues to initiate opportunistic interactions in face-to-face settings. Previous research observed that visible smiling frequently accompanies greeting utterances, facilitating conversation openings~\cite{kendon1973greetings}. Additionally, individuals may adjust their body positions during moments of co-presence to ease the initiation process, allowing them to reset spatial and territorial orientations for a smoother approach~\cite{goodwin1981conversational, mortensen2014moving}. In another study, Keevallik analyzed a full-day recording of 11 young people working together, finding that factors such as posture, gaze, and the bodily orientation of fellow workers contribute to attracting responses from others~\cite{keevallik2018sequence}.

In short, the above literature reveals that multiple elements, both verbal and non-verbal, contribute to the successful initiation of opportunistic interactions during face-to-face encounters. While multiple signals can jointly influence a person's perceived availability and willingness to engage, the duration of this assessment is typically as brief as a few seconds~\cite{pillet2018begin}. The initiation of a face-to-face interaction is generally regarded as a single, cohesive, and brief process.

\subsection{Video-Mediated Opportunistic Interactions in Professional Contexts}\label{opportunistic interactions_Online}
Over recent decades, advancements in computer-mediated communication (CMC) tools have profoundly reshaped communication in professional contexts, transitioning from solely co-located, face-to-face interactions to more flexible, frequently remote environments~\cite{fish1992evaluating, zhao2009and, egido1988video}. The shift was accelerated by the COVID-19 pandemic, which prompted many companies to adopt fully remote work models through team-based video conferencing~\cite{fana2020telework, bennett2021videoconference, hacker2020virtually}. Despite this flexibility, remote settings greatly limit opportunistic interactions that naturally occur in physical workspaces~\cite{fay2011informal, bleakley2022bridging}. Previous literature attributes this reduction in CMC-based opportunistic interactions to two main factors: the lack of spatial relationships in video conferencing—similar to hallways in physical offices that facilitate casual encounters~\cite{rudnicka2022end, isaacs1993video}—and the absence of rich cues prevalent in face-to-face interactions~\cite{grayson2003you, fish1992evaluating, doherty1997face}.

CSCW researchers have long investigated ways to foster a sense of co-location in distributed video conferencing (e.g., ~\cite{hauber2006spatiality, raskar1998office, gibbs1999teleport, vertegaal1999gaze, benford2001collaborative, elrod1992liveboard}). For example, early work by Fish et al. attempted to simulate physical proximity by placing human-scale screens in workspaces on different floors, encouraging workers in separate locations to engage in spontaneous conversations~\cite{fish1990videowindow}. Later, CRUISER~\cite{fish1992evaluating, root1988design} and Montage~\cite{tang1994supporting, tang1994montage} extended this concept to desktop-based video software. CRUISER was inspired by the concept of roaming a hallway and meeting people without prior planning. It facilitated both intentional and system-generated spontaneous video connections between users. Montage, on the other hand, introduced a gradual appearance and fading effect for the other person's image, simulating movement closer to or farther away to mimic interactions in physical spaces. 

While these early explorations were not directly integrated into later video conferencing systems, the concept of rendering proximity has become widely adopted. More recent commercial platforms such as Gather.town~\footnote{Gather.town: \url{https://www.gather.town/}} use a shared 2D spatial layout where distributed users can see and approach each other. When users come into close proximity, a traditional video chat interface appears, allowing them to initiate spontaneous interactions. A recent study by Sanchez et al. found that such features indeed enhanced users' awareness of each other, making it easier to determine when to interact, interrupt, or initiate a conversation. However, Gather.town users still reported insufficient information on whether others were actually interruptible for a conversation, essentially due to the lack of non-verbal cues within the 2D spatial layout interfaces~\cite{palos2023exploring}.

It should be noted that non-verbal cues, such as facial expressions, gaze, and gestures, are widely utilized during video conferencing~\cite{nguyen2009more}. Studies have shown that gestures and facial expressions can convey subtle emotions~\cite{isaacs1993video}, whereas gaze can foster trust when well-perceived through video streaming~\cite{nguyen2007multiview}. However, unlike in face-to-face settings, these non-verbal exchanges in video-mediated interactions generally come into play only after a conversation has begun. The reliance on video cameras and 2D screens also limits the effectiveness of non-verbal cues for initiating opportunistic interactions~\cite{j2013comparative}.  Thus, although considerable research has sought to enhance opportunistic interactions in video conferencing, it remains a challenging task.

\subsection{Opportunistic Interactions in Social VR}\label{opportunistic interactions_SVR}
While VR has long been a key area of research in HCI~\cite{burdea2024virtual}, its use in social contexts has only recently gained attention, driven by the rise of affordable commercial VR hardware, such as the Meta Quest\footnote{Meta Quest: \url{https://www.meta.com/quest/}} and HTC Vive\footnote{HTC Vive VR Headset: \url{https://www.vive.com/us/}}. On social VR platforms, multiple users can interact with each other in 3D-rendered virtual spaces, accessible either via head-mounted displays (HMDs), or desktop setups using keyboard and mouse devices~\cite{mcveigh2019shaping, mcveigh2018s}.

Research has identified three major technical features unique to social interactions in VR. First, VR offers full-body avatars and real-time gesture capabilities within high-fidelity, immersive 3D environments~\cite{freeman2021body}. Second, it supports rich spatial and temporal experiences, fostering emotional connections similar to those in face-to-face interactions~\cite{chang2024social}. Third, VR enables both verbal and embodied communication, allowing for a broad range of social activities~\cite{mcveigh2019shaping, moustafa2018longitudinal, sra2018your, zamanifard2019togetherness, maloney2020talking, zhang2025speechcap}. Consequently, unlike video conferencing, which is widely used in professional workplace settings, VR has been naturally adopted for more casual social gatherings~\cite{maloney2020falling}. In a recent study, Freeman et al. interviewed 30 social VR users and found that their primary goal was to meet new people rather than maintain existing relationships~\cite{freeman2021hugging}. Further research by Maloney et al. observed how informal encounters in VR led to the initiation of opportunistic interactions. They found that a wide range of non-verbal signals were used to initiate connections, from simple nods and waves by avatars to more dynamic and sometimes ``wild'' gestures, such as dancing, kissing, pushing, bumping, and even blocking others' view to capture attention~\cite{maloney2020talking}. Many of these strategies would be uncommon or even considered inappropriate in face-to-face interactions between people meeting for the first time. Nevertheless, these ``unconventional'' behaviors are far more tolerated in the social VR context.

More recently, a growing number of professional events have been hosted in social VR, triggered by the COVID-19 pandemic and continuing beyond it (e.g., ~\cite{ahn2021ieeevr2020, moreira2022toward, mulders2021academic}). 
\edited{For example, IEEE VR 2021 allowed attendees to immerse themselves in the Virbela space to watch talks and interact with each other ~\cite{moreira2022toward}. Another academic event, VRARBB@SocialVR, leveraged AltspaceVR for topic-oriented discussions over six weeks of conference time ~\cite{mulders2021academic}. While participants in both events reported VR to be a suitable medium for professional gatherings, limited information is available on how opportunistic interactions actually unfolded in these environments.} 
This body of literature prompted us to ask the following research question (RQ):

\textit{\textbf{RQ.}}  How do people initiate opportunistic interactions at professional events in social VR?

\section{Method}

We investigate the above RQ through a qualitative study involving 16 individuals. With the participants' consent, we shadowed them to observe their opportunistic interactions with others at 23 distinct professional events across 4 social VR platforms. Each event lasted between one and two hours, with total attendance ranging from twenty to fifty. We also conducted in-depth interviews with each participant after their events. Following the theoretical sampling process suggested by grounded theory ~\cite{glaser1978theoretical}, we iterated through generating codes from collected data and reviewing and elaborating these codes by collecting more data. We stopped the data collection once all core variables reached saturation. The remainder of this section details our recruitment strategy, the types of data being collected, and the data analysis approach. \edited{The study was reviewed and approved by the University Institutional Review Board (IRB).}

\subsection{Recruitment Strategy}

The majority of our recruitment took place through the dissemination of digital flyers via two channels: internal mailing lists within the researcher’s institutions and VR-relevant public interest groups on social media. Qualified participants must be a) at least 18 years of age, b) comfortable speaking with the researcher in English, and c) have already signed up for one or more VR-based professional events to attend prior to enrolling in our research. In addition, we purposefully looked for individuals with various backgrounds in terms of the role(s) they had acted in at the target events, the platform(s) they had used, and their overall experience with social VR. This strategy enabled us to remain sensitive to the potential similarities as well as differences across them. Our final sample consisted of 16 individuals who satisfied all the above criteria. Table~\ref{table:1} details the background of each participant.

\subsection{\edited{Event Contextualization}}
\edited{All 23 professional events in this study were public, stand-alone gatherings. Unlike academic conferences such as IEEE VR~\cite{ahn2021ieeevr2020, moreira2022toward}, which host a series of VR talks, panels, and discussions over multiple days with the same group of participants, the events we attended were one-off gatherings, primarily aimed at fostering networking opportunities among professionals within a time frame of 60 to 120 minutes. Attendees at these events usually shared a common interest, but they often did not know each other beforehand.}

\edited{A typical event began with an opening, often a short talk or panel discussion. Depending on the event, the themes of these openings were usually tailored to the specific professions involved. For example, a professional event focused on technology might begin with a speech on blockchain communities and their ongoing investment activities, whereas an event centered on art and design could start with a panel discussion about the career potential of digital art. These introductions were generally brief though, as they mainly served to provide context for the event and to set the stage for the networking activities that followed.}

\edited{Each event allocated substantial time for networking, which typically occurred in one of two formats. Some events featured unstructured networking sessions, where attendees could meet and interact with other professionals organically. A few events framed the networking component as a tour of a virtual space, where attendees were encouraged to engage in open conversations during the tour. Alternatively, they were free to leave and rejoin at their convenience.}

\edited{The events we attended generally fell into the following professional communities: art and design, where events were mainly hosted by artists or organizations that promote digital art; technology, where events were led by technologists, hobbyists, and entrepreneurs, such as those from gaming companies; and education, where events were organized by educators or librarians, for example, those interested in extending classroom activities into VR.}

\vspace{2ex}

\noindent\begin{minipage}{\linewidth}
\centering

\captionof{table}{Background information of all participants.}\vspace{-2ex}\resizebox{\linewidth}{!}{%
\begin{tblr}{
  column{1} = {c},
  column{2} = {c},
  column{4} = {c},
  column{5} = {c},
  column{7} = {c},
  column{8} = {c},
  column{10} = {c},
  column{11} = {c},
  column{12} = {c},
  column{13} = {c},
  cell{1}{1} = {r=2}{},
  cell{1}{2} = {r=2}{},
  cell{1}{3} = {r=2}{},
  cell{1}{4} = {c=2}{},
  cell{1}{7} = {c=2}{},
  cell{1}{10} = {r=2}{},
  cell{1}{11} = {r=2}{},
  cell{1}{12} = {r=2}{},
  cell{1}{13} = {r=2}{},
  cell{3}{6} = {c},
  cell{3}{9} = {c},
  cell{4}{6} = {c},
  cell{4}{9} = {c},
  cell{5}{6} = {c},
  cell{5}{9} = {c},
  cell{6}{6} = {c},
  cell{6}{9} = {c},
  cell{7}{6} = {c},
  cell{7}{9} = {c},
  cell{8}{6} = {c},
  cell{8}{9} = {c},
  cell{9}{6} = {c},
  cell{9}{9} = {c},
  cell{10}{6} = {c},
  cell{10}{9} = {c},
  cell{11}{6} = {c},
  cell{11}{9} = {c},
  cell{12}{6} = {c},
  cell{12}{9} = {c},
  cell{13}{6} = {c},
  cell{13}{9} = {c},
  cell{14}{6} = {c},
  cell{14}{9} = {c},
  cell{15}{6} = {c},
  cell{15}{9} = {c},
  cell{16}{6} = {c},
  cell{16}{9} = {c},
  cell{17}{6} = {c},
  cell{17}{9} = {c},
  cell{18}{6} = {c},
  cell{18}{9} = {c},
  cell{19}{6} = {c},
  cell{19}{9} = {c},
  cell{20}{6} = {c},
  cell{20}{9} = {c},
  cell{21}{6} = {c},
  cell{21}{9} = {c},
  cell{22}{6} = {c},
  cell{22}{9} = {c},
  cell{23}{6} = {c},
  cell{23}{9} = {c},
  cell{24}{6} = {c},
  cell{24}{9} = {c},
  cell{25}{6} = {c},
  cell{25}{9} = {c},
  hline{1,3,26} = {-}{},
  hline{2} = {4-5,7-8}{},
}
{\textbf{Event }\\\textbf{ID}} & {\textbf{Participant }\\\textbf{ID}} &  & \textbf{Observed Events} &                   &  & \textbf{Other Distinct Historical Events} &                                   &  & \textbf{Professional Community} & \textbf{Age} & \textbf{Gender} & {\textbf{Frequency of Attending }\\\textbf{Social VR Events}} \\
                                 &                                      &  & \textbf{Role}            & \textbf{Platform} &  & \textbf{Role}              & \textbf{Platform}                 &  &                    &              &                 &                                                               \\
1                                & P1                                   &  & Attendee                 & VRChat            &  & Attendee                   & VRChat, AltSpace                  &  & Art and design     & 25-34        & Female          & Several times a year                                          \\
2                                & P1                                   &  & Attendee                 & AltSpace          &  &                            &                                   &  &                    &              &                 &                                                               \\
3                                & P2                                   &  & Attendee                 & VRChat            &  & Attendee                   & VRChat, AltSpace, Venu            &  & Tech development   & 25-34        & Male            & Several times a year                                          \\
4                                & P2                                   &  & Attendee                 & Venu              &  &                            &                                   &  &                    &              &                 &                                                               \\
5                                & P2                                   &  & Attendee                 & Venu              &  &                            &                                   &  &                    &              &                 &                                                               \\
6                                & P3                                   &  & Attendee                 & AltSpace          &  & Attendee                   & VRChat                           &  & Education           & 18-24        & Female          & Several times a year                                          \\
7                                & P4                                   &  & Attendee                 & VRChat            &  & Attendee                   & VRChat                            &  & Art and design     & 25-34        & Female          & Several times a month                                         \\
8                                & P4                                   &  & Attendee                 & VRChat            &  &                            &                                   &  &                    &              &                 &                                                               \\
9                                & P5                                   &  & Attendee                 & AltSpace          &  & Attendee                   & VRChat, Horizons                  &  & Education           & 35-44        & Male            & Several times a year                                          \\
10                               & P6                                   &  & Attendee                 & AltSpace          &  & Attendee                   & VRChat                            &  & Education           & 18-24        & Male            & Several times a year                                          \\
11                               & P7                                   &  & Attendee                 & Venu              &  & Attendee, Host             & AltSpace, VRChat        &  & Tech development   & 35-44        & Male            & Several times a month                                         \\
12                               & P8                                   &  & Attendee                 & Venu              &  & Attendee, Host             & AltSpace                          &  & Tech development   & 25-34        & Male            & Several times a month                                         \\
13                               & P9                                   &  & Attendee                 & Engage            &  & Attendee                   & AltSpace                          &  & Education          & 45-54        & Female          & Several times a year                                          \\
14                               & P10                                  &  & Attendee                 & Engage            &  & Attendee                   & VRChat, AltSpace                  &  & Tech development   & 25-34        & Male            & Several times a month                                         \\
15                               & P11                                  &  & Attendee                 & Engage            &  & Attendee                   & VRChat                            &  & Tech development   & 25-34        & Male            & Several times a year                                          \\
16                               & P12                                  &  & Host                     & AltSpace          &  & Attendee                   & AltSpace, VRChat, Engage          &  & Tech development   & 18-24        & Male            & Several times a week                                          \\
17                               & P12                                  &  & Host                     & VRChat            &  &                            &                                   &  &                    &              &                 &                                                               \\
18                               & P13                                  &  & Host                     & Venu              &  & Attendee                   & VRChat                            &  & Tech development   & 25-34        & Male            & Several times a week                                          \\
19                               & P14                                  &  & Host                     & Venu              &  & Attendee                   & VRChat                            &  & Tech development   & 25-34        & Male            & Daily                                                         \\
20                               & P15                                  &  & Host                     & VRChat            &  & Attendee                   & VRChat, RecRoom                   &  & Art and design     & 25-34        & Male            & Several times a week                                          \\
21                               & P15                                  &  & Host                     & VRChat            &  &                            &                                   &  &                    &              &                 &                                                               \\
22                               & P16                                  &  & Host                     & AltSpace          &  & Host                       & Engage                            &  & Education           & 45-54        & Female          & Several times a week                                          \\
23                               & P16                                  &  & Host                     & AltSpace          &  &                            &                                   &  &                    &              &                 &                                                               
\end{tblr}
\label{table:1}
}
\end{minipage}

\subsection{Observation and Interview Protocol}\label{protocol}

We developed two semi-structured protocols to guide our observations and interviews with each participant, respectively.

\textit{Observation Protocol \minor{ and Consent Approach}.} With participant consent, we observed one or more professional events they attended within their identified professional community. Before each observed event, the researcher attended several additional events with the participant, upon invitation, to become familiar with the participant’s professional community. 

During the observed event, the researcher maintained an appropriate distance from the participant, refraining from initiating any verbal or non-verbal interactions with the participant or others unless approached. The observation notes collected from each event document: a) interactions between the participant and others, especially the participant’s attempts at initiating opportunistic interactions; b) behaviors that the researcher deemed worth discussing in post-event interviews; and c) the researcher’s own experiences as an observer. When an observation ended, the researcher sent her notes to the participant for review and removed information that the participant preferred not to be recorded, if any.

\minor{Other individuals who may have interacted with the participant were not individually consented. Instead, we implemented proxy consent, in which a third party acts as an intermediary to inform others about the research activity~\cite{humphreys2015reflections},  to encourage transparency and reduce risk. While not equivalent to directly informed consent, proxy consent is often used in observational research where obtaining direct consent from all involved is impractical. In particular,  we informed each event’s host of our observation plans in advance. These hosts, typically recognized as organizers or moderators in the community, were given discretion to decide whether to verbally announce the researcher’s presence and purpose at the beginning of their event (e.g., clarifying that the researcher was not collecting any personally identifiable information).  Correspondingly, the researcher entered events with a virtual nametag diplayed above her avatar, identifying her role as a researcher to all attendees at the event. }

\textit{Interview Protocol}. The same researcher also interviewed each participant within 24 hours following the event at which they were observed. All interviews took place via Zoom, with each lasting around one hour. Participants were prompted to share their reflections on three main topics: a) their experience at the event they attended; b) their interactions with others during the event, especially their attempts to initiate conversations, whether successful or not; and c) their professional interactions via social VR but outside the observed event. When participants described a specific moment of interaction at the event, the researcher encouraged them to recall the purpose of that interaction, how it was initiated (or failed), and their experience during that interaction. All interviews were audio recorded and then transcribed verbatim for analytical purposes.

\subsection{Data Analysis}

We followed an inductive process to analyze all data obtained from interviews and observations. To start, the first and last authors of this paper independently reviewed all data. Each researcher developed an initial list of codes to comprehensively cover a randomly selected subset of data. They then reviewed these codes together, made revisions, and carried insights from these discussions to independently code the next subset. As the two researchers converged on a shared list of codes and concepts, the third author joined to review all materials generated so far. This auditing step helped ensure the rigor and credibility of the analysis. We repeated the process until reaching saturation. We then iterated between examining all emergent concepts in the context of our research and reviewing related literature to uncover more latent relationships among those codes and concepts. 

By the end of this analytical process, we identified the notion of verisimilitude as a focal point connecting different categories of concepts, providing an organic explanation of how participants initiated opportunistic interactions in response to our RQ. We present more details about our findings in the next section. During this presentation, we use the participant ID to indicate the source of each referenced observation and/or interview. Direct quotations from participants are italicized for easy differentiation from other text.

\section{Findings}
The bulk of our data highlighted the role of verisimilitude---the extent to which social VR preserves the real-world social meanings of various cues as perceived by its users---in shaping people's initiation of their opportunistic interaction with others. All participants underscored the importance of deliberately considering the verisimilitude of non-verbal cues. They concluded, based on prior experiences in social VR, that \textit{``it's better to start verbal exchanges only after you are sure the other person is up for it.''} Directly talking to someone would \textit{``not [be] a smart move''} due to the often unstable audio quality at social VR events involving a large number of attendees. Issues such as \textit{``audio keeps clipping in and out''} and \textit{``noises come in from someone's physical environment [while using VR]''} were mentioned in the interview response of every participant, along with the occasional problem that \textit{``the volume of speaking may not work right by default.''} Non-verbal cues serve as their strategic tools, allowing people to draw others into opportunistic interactions while sidestepping the social awkwardness of abrupt speaking.  

More specifically, our data revealed how participants interpreted and acted upon the verisimilitude of non-verbal cues at three distinct steps toward opportunistic interactions in social VR: recognizing others' availability (section~\ref{recognizing availability}), gaining their attention (section~\ref{gaining attention}), and breaking the ice (section~\ref{breaking ice}). We use these steps to structure the presentation of our main findings, followed by participants' overall reflections regarding the challenges of acquiring established interaction protocols in social VR (section~\ref{acquiring protocols}). 

\subsection{Recognizing the Availability of Others}\label{recognizing availability}
Participants considered the successful recognition of another person's availability to be the starting point for opportunistic interaction in social VR, just as it is in the real world. They attempted to infer by reading the non-verbal cues that others' avatars have displayed: is the person behind this avatar occupied at the moment, or are they open to a chat? 

Eye gaze and body movement were reported by every individual in our sample as their most frequently used non-verbal cues for availability assessment. An avatar gazing toward someone intuitively indicates that the person behind it is willing to be approached. An avatar wandering around signals that the person is looking for opportunities to connect with others. While all participants tended to draw the above inferences based on corresponding experiences at real-world events, many reported reconsidering the verisimilitude of these intuitive cues after witnessing counterexamples. Our data analysis identified common characteristics among the counterexamples discussed by participants (section~\ref{blurred boundary}), as well as their coping strategies (section~\ref{non occupied}). 

\subsubsection{A blurred boundary between actions and reactions}\label{blurred boundary}
The visual design of an avatar's eyes and body, across all social VR platforms examined in the current research, closely mimics human appearance in the real world. This design contributed to a sense of high verisimilitude perceived by our participants, suggesting that the motion of eyes and body in VR would also convey similar social meanings to those in real world.

For the purpose of availability assessment, participants have learned from their everyday experience that gazing and wandering are \textit{``agentic actions.''} These cues indicate that a person is consciously opening themselves up to further interactions with others. The quotation from P9 illustrates how cues-as-actions matter for opportunistic interaction: 

\begin{quote}
    \textit{``When you make an action, I understand your intention better. When you gaze or walk [toward] my direction, I understand you are looking for someone to talk. I won't come to you [to start an interaction] if I realize you are just reacting to something.''} [P9]
\end{quote}

Interestingly, the inherent meaning of gazing-as-actions is often disrupted in social VR. The rules governing an avatar's eye gaze vary considerably across platforms. To name a few examples, AltspaceVR codes its avatars such that new sound in the virtual environment will trigger the avatar to gaze toward the direction of the sound; in Engage, avatars' eyes gaze slightly and blink occasionally without being controlled by their human users; in VRChat, an avatar's eyes will, by default, look around, blink randomly, and sometimes fixate on others nearby, while it also allows for actual gaze tracking for users with compatible headsets. %
None of our participants fully articulated all of these differences during the study, regardless of their prior experience with social VR.

The above rules largely convert gazing in social VR into reactions determined by the technical setup of the virtual world, rather than actions controlled by the people behind their avatars. Many participants had encountered incidents suggesting that virtual gaze may differ from its real-world counterparts in important ways. While this insight is valuable, it did not provide further guidance on how to calibrate or restore the social meaning of gazing in VR. As shared by P1 and P16: 

\begin{quote}
    \textit{``Gazing in VR can be confusing. For example, there was this person at today's event. She was, like, looking at me. I felt we even made eye contact. But later, she passed right through me to someone behind me. It wasn't meant for me, even though her eyes were looking in my direction. I didn't know if it was accidental. I was so confused, but it's okay. I tell myself `I'm just assuming it doesn't mean anything.' ''} [P1]
\end{quote}

\begin{quote}
    \textit{``Humans make paintings where the eyes follow us. People say, ‘eyes are the window to the soul.' There's a lot of focus on eyes. It's true that eyes are very important. But in VR, I don't look at the eyes anymore. I don't even care what the eyes do. I just use the eye gaze to double-check where their face may be attentive to, and that's it. It would be great if eyes in VR would indicate more, but they don't.''} [P16]
\end{quote}

Similar confusions also appeared to challenge some participants' belief about wandering-as-actions in social VR. The more experience participants gained with the VR technology, the more they became aware that an avatar's body movements may reflect not only a person's actions at the virtual event but also their actions in the physical environment. 

Specifically, participants recalled instances where they attended to things happening in their physical space, such as \textit{``the cat came over to play with me''} or \textit{``my family needed me for something,''} while attending events in social VR. In these situations, the person's avatar often wandered in the virtual reality as a reflection of actions intentionally performed in the physical reality. However, this intention does not transfer across worlds, making the distinction between action and reaction unclear for others who witnessed the avatar's body movement at the moment. 

\subsubsection{From ``not occupied'' to ``ready for a chat.''}\label{non occupied}
We asked participants about alternative means they had developed, if any, to recognize others' availability in social VR. Not all participants had an answer to this question. Nevertheless, the responses from those who did converge on one common strategy: it would be more effective to confirm a person is \textit{``ready for a chat,''} rather than merely \textit{``not occupied.''} Gazing and wandering are intuitive cues for the latter but not necessarily the former.  

Most social VR platforms today offer visual indicators of a user's speaking status. When the system detects ongoing speech from the person behind an avatar, the avatar's mouth repeatedly switches between open and closed. Meanwhile, a speech icon often appears near the avatar's head, and it goes off as soon as the speech ends. Our participants paid close attention to this set of cues. %
By doing so, they managed to gain a good sense of who else was in an (inter)action mode already and, more importantly, when they had finished their last chat and thus became available for the next one. As explained by P10:

\begin{quote}
    \textit{``I observe those [avatars] around me and wait until there's a moment of pause [in their ongoing interactions]. [I] make sure no speech icons are still on around their heads. That would be my opportunity to interject if I've played everything right. I guess the green icon (which refers to the speech icon) is kind of like a check mark. When the green stops flashing, it means I can initiate something because they are ready for another chat again, without me being rude.''} [P10]
\end{quote}

\textbf{To summarize}, the above findings suggest that people often leveraged the avatar's eye gaze and body movement for availability assessment in social VR. Yet, the technical setup of virtual platforms can diminish the verisimilitude of gazing and wandering, despite their lifelike appearance. Some participants turned to selected alternatives, such as a combination of the avatar's mouth movement and speech icon, for supplementary information. While these additional cues increased people's confidence in recognizing others' availability for the next opportunistic interaction, they restricted interactions only to those who had attended other chats already. 

\subsection{Capturing the Attention of Others}\label{gaining attention}

Our observation notes documented numerous instances where participants leveraged various non-verbal cues to capture others' attention, in the moments between recognizing others' availability and speaking to them. During interviews, all participants emphasized that, in contrast to the real world, gazing and other cues in the virtual environment provide hints about a person opening their attention to someone else but \textit{``(can)not specify where exactly that attention will go.''} Such a contrast makes attention-grabbing a crucial step for successful opportunistic interaction in social VR.  

Clapping hands, making positive facial expressions, and reducing the proximity between avatars appeared as popular cues among many else to gain others' attention in social VR. These popular cues share two qualities valued by our participants: they convey the social meaning of someone being \textit{``welcoming''} or \textit{``receptive to others''} with high verisimilitude, and they are visually appealing. Moreover, we found that participants did not make impulsive decisions about which specific cue(s) to activate for attention-grabbing at a given moment. Instead, they learned the constraints of each cue through trials-and-errors in social VR, using that insight to improve their initiation of opportunistic interactions with others. Below, we detail three specific constraints discussed by our participants, each affecting their use of certain cues over others (sections~\ref{instant vs delay}-\ref{flexibility vs social pressure}).

\subsubsection{Instant vs. delayed effect}\label{instant vs delay}
As a person's attention can shift from one spot to another at any time, all participants highlighted the importance of gaining others' attention at the right timing. Many elaborated on this point in the context of issuing handclaps or positive facial expressions---the latter cue is often conveyed via the clicking of relevant emojis in today's social VR platforms. P6 described why the timing of handclapping or emoji sending was critical, which aligns with the reflections made by many others:

\begin{quote}
    \textit{``A good way to get other people's attention, [while acting] naturally, is by giving compliments or affirmation, at the time as they speak and we are standing nearby. The clap and the emoji, they are especially useful in VR to express, ‘Oh, I'm here for you.' …In real-life settings, if someone smiles or claps [in response] to a speaker, I think it's the same thing in VR. Also, in real life, you will need to be quick, right? The other person's status [of attention] can change quickly, or it only lasts a short time. So [when using] the Altspace emoji, I have to actively tap it to catch the timing.''} [P6]
\end{quote}

In practice, not all individuals can make fluent use of handclaps and emojis. According to our observations, a person needs to be fairly sophisticated in using their hand controller so that their avatar could perform the handclapping action smoothly.
In the case of issuing emojis, people must pull out a list in the virtual environment and select their desired expression, as all commercial VR headsets today lack facial tracking capabilities.  %

Our participants were often \textit{``thrown off''} by the time delay in getting their avatar's handclap or emotion expression work properly. Even for those who self-identified as experienced VR users, it was difficult to guarantee the instant activation of these cues on any given social VR platform and at all times. As said by P2 and P6: 

\begin{quote}
    \textit{``Hand gestures, like clapping, can send a positive message [to the person you hope to interact with more]. This is helpful for building a conversation. It shows you look forward to more interactions. But these types of signals happen in milliseconds. The avatar usually acts with a lag, and the opportunity passes . It's frustrating.''} [P2]
\end{quote}

\begin{quote}
    \textit{``I did use the emoji [in social VR], but it was slow. I have to find the right one in my search bar and then click it. It's not as immediate, so users must have very fast hands. It's a bit annoying because, to start the interaction with that person, you do not have a lot of time. I need to make the interaction right there and in the moment, or people will lose interest quickly. But in VR, everything is like half a second slower for me. Maybe if I use it more, I will be faster, [but] I'm not sure.''} [P6]
\end{quote}

Reducing the time delay between the actions issued by real people and the actions displayed by their avatars should enhance the verisimilitude of relevant cues in social VR. Our data collection with participants using Venu have provided positive support for this notion. In particular, Venu hardcoded two short paths to trigger the handclapping of an avatar: clapping the two hand controllers or pressing ``1'' on a keyboard. At all the events we observed on this platform, handclapping appeared as the most frequently used cue for participants to guide the attention of others. They also reported in the interviews that handclapping on Venu \textit{``was a really great feature for some easy non-verbal communication.''}

\subsubsection{Multimodal vs. visual sensing}\label{multimodal}
Access to visual information in social VR (and audio, although many would by default put themselves on mute when not speaking) allows people to leverage various non-verbal cues for opportunistic interaction. That said, multiple participants reported a \textit{``feeling of dissonance''} when interpreting the visual distance between their avatar and others: reduced proximity sometimes did not prompt an interpersonal exchange of attention, which differs from people's real world experiences. %
As explained by P3:

\begin{quote}
    \textit{``In everyday life, people adjust their distance with others. We adjust our personal space to [indicate that we] welcome others into our space. I do the same in VR, but I don't actually know how close I am to other people. I can still tell if we are far away from each other or I'm moving closer to you. But something [about VR] makes me desensitized to what that distance means [for social interactions].''} [P3]
\end{quote}

Although our participants did not name a definite reason for this dissonance, many suspected that the lack of multimodal sensing disrupts the verisimilitude of proximity in social VR. As they pointed out, the translation from shorter physical proximity to greater social attentiveness is processed beyond just the visual perception of a person's image getting larger. social VR today preserves the visual aspect of this translation, but it \textit{``leaves out something else.''}  P5 and P11 shared their understanding of what that \textit{``something else''} is:

\begin{quote}
    \textit{``When I approach [in the real world], you're just going to notice that without looking. You sense the combination of the change in light, temperature, and the way the air feels on your skin. There are so many subtle cues that you actually sense in real reality and not anymore in VR. So, in VR, say I approach from behind or out of your vision, especially if I'm mute, you won't be able to notice me. But our social exchanges rely on the combination of all these cues you pay attention to.''} [P5]
\end{quote}

\begin{quote}
    \textit{``When you walk closer to a real person, you can feel the vibe that they are gonna pick up on your presence. And you progressively add to it. It's hard to describe because it's just like second nature, but maybe it starts with, like, I kind of raise my chest up towards them a bit, and then I use my eyes, and my shadow moves around or something. Usually that's how people determine, ‘okay I can engage this person.' All these subtleties are missing in VR.''} [P11]
\end{quote}

\subsubsection{Granting flexibility vs. imposing social pressure}\label{flexibility vs social pressure}
Three participants (P2, P12, and P9) envisioned that future social VR platforms could add \textit{``virtual business cards''} as an embedded feature, especially for people using the platform to attend professional events. This function mimics similar etiquette established in real-world conferences. By pressing a short key assigned to trigger the card exchange, social VR users would be able to gain attention from one another without worrying about constraints discussed in above sections.

We sought the other participants' reactions to the \textit{``virtual business cards''} idea. While some felt this card exchange would really help \textit{``seal the deal and encourage more interaction to happen,''} others disliked it. For them, one unique value of social VR is the flexibility built into interpersonal exchanges. People can make themselves \textit{``visible but also anonymous enough''} in front of others.  They can more easily avoid the social pressure of \textit{``feeling obligated to respond to others,''} which is common in face-to-face interactions. However, the etiquette of card exchange, if introduced into the virtual environment, might impose too much social pressure on the recipient. It would not allow an equal level of flexibility to that existing in the exchange of other attention-grabbing cues, such as facial expressions via emojis and reduced proximity between avatars.

\textbf{Together}, our data shows that hand gestures, facial expressions through emojis, and proximity adjustment are all useful cues for people to capture others' attentions in social VR. Participants discussed three constraints that could disrupt the verisimilitude of these cues. As our analysis has demonstrated, each cue was perceived as sensitive to some of these constraints but resilient to others. To mitigate the potential loss of social meanings, P6 reported that he followed a \textit{``give and take protocol:''} when there was a person he really wished to speak with, she would first observe the attention-grabbing cue that person used most frequently, then he adopted the same cue to initiate his interaction with them. Notably, nobody else reported using the same strategy, probably due to the effort required to implement it. Issuing multiple cues simultaneously appeared as a much more common strategy used by most participants to gain others' attention for opportunistic interaction.

\subsection{``Breaking the Ice'' with Others}\label{breaking ice}
The exchange of attention in real-world meetups often naturally leads to small talks between two people---whether it's a brief chat about the weather, local happenings, or something else---which eventually transitions into more meaningful conversation. A substantial part of our data reveals how this ice breaking step occurs at professional events within social VR. Not surprisingly, all participants found it more challenging to really start a conversation in the virtual environment compared to the real world. We kept hearing from different individuals that \textit{``VR doesn't offer many clues about who is more friendly or approachable [than others]''} and \textit{``Looking at people standing in VR, we can't really tell if they are having a good or bad day, [making it] hard to start a conversation from there.''} In essence, people either lack non-verbal cues to craft effective ice-breakers, or the accessible cues convey insufficient verisimilitude to indicate the common ground between individuals. 

While there were no sustainable solutions that emerged from our data, we did observe two sets of actions taken by participants as their current best practices for kicking off conversations in social VR. In one set of actions, participants leveraged features of a person's customized avatar to infer the likes and dislikes of that person, and then crafted topics for their small talk from there (section~\ref{appearance}). The other set of actions was observed exclusively, yet frequently, with the events' hosts: these individuals actively created talking points for others, enabling attendees at the event to engage in group interactions easily, then break off into one-on-one chats if preferred (section~\ref{warm-up}). We elaborate on both sets of actions in the following paragraphs. 

\subsubsection{Avatar appearance as an indicator of the person's interests}\label{appearance}
All social VR platforms involved in the current research allow users to customize their avatars, although some provide richer options than others. For our participants who used these platforms for professional events, having a customized avatar design was essential to bringing their real-world identity into the virtual world. They wanted their avatars to feature certain demographics, especially their race and gender, or dressing styles that represented how they would typically appear at daily work in the real world. As underscored by P12 and P7:

\begin{quote}
    \textit{``When I go to these types of professional events in VR, I make my avatar look like me, especially my hair and glasses and other accessories, because, well, we want to create connections [with others in the community], and that may extend outside of VR as well. Professionalism is important. Platforms like VRChat and other ones have been offering more customizations over the years. [With that], I've tried to make my avatar look closer and closer to what I might wear in physical reality.''} [P12]
\end{quote}

\begin{quote}
    \textit{``I wanted to make sure I wasn't just the default avatar. I want to convey my style of dress, and also my race as far as I can in there. I'm trying to make it as close to the real life as possible. Because when I go to these professional meetups, I'm open to connecting with these people in real life later. I don't want to appear too different from my avatar in real life. If it was not a professional event, it wouldn't be a big deal. I'm hoping this will become a real human relationship outside of avatars.''} [P7]
\end{quote}

The presumed connection between an avatar's appearance and the characteristics of the person behind it directed some participants to initiate small talk with people whose avatars included common features with theirs. These participants reported feeling more confident and comfortable breaking the ice this way, as their avatars signaled topics that were likely to align with both parties' shared interests. As explained by P13 and several others: 

\begin{quote}
    \textit{``I assume that the way we dress or present ourselves determines how we make inferences about each other [before talking for the first time]. From the way my avatar is dressed and behaves, people can see this is me, and they can sense what I like or what I'm good at. Then, some can identify with me, and they choose the right things to talk about with me.''} [P14]
\end{quote}

\subsubsection{Introductory chats and warm-up activities initiated by the hosts}\label{warm-up}

We noticed from the interviews that participants who actively leveraged the avatar appearance for ice-breaking were usually individuals who described themselves as \textit{``outgoing''} or \textit{``talkative.''} The majority of our participants often did not feel fully comfortable self-initiating small talk with others, even at times when they were already in a close distance to others and using some attention-grabbing cues, as discussed in the previous sections. Across the events we observed, it was not uncommon to see small groups of people standing quietly together until some eventually decided to speak up (or left). 

Scaffoldings provided by event hosts appeared to be a more reliable means toward successful ice-breaking among individuals. At every event involved in the current research, we witnessed such practices and heard relevant reflections during interviews. Much like in real-world events, hosts in social VR considered creating talking points for others as their \textit{``tacit duty.''} For those we observed hosting, surveying the entire virtual space and initiating small talk with individuals who \textit{``aren't quite as quick to hop into a conversation''} were the actions they consistently repeated from time to time. Their role as host made it natural for them to initiate conversation on topics relevant to event management, such as asking others about their experience so far, or suggesting places to explore within the virtual event hall. These exchanges set opportunities for people to hear one another's responses, easing them into further interactions after the host moved on. 

More interestingly, hosts invented unique group activities as \textit{``social warm-ups,''} relaxing people and helping them foster a sense of connection for easier interaction. We received many positive remarks from attendees who had participated in such activities under their host's guidance. The most intriguing aspect of these activities, according to our participants, was the quickly spreading joyfulness among individuals as they engaged collectively. 

One example documented in our observation notes was from the event hosted by P13 on Venu. At the moment captured in this note, P13 approached a group of people who were quietly standing together. He introduced himself as the event host and invited everyone to clap their avatars' hands with him. In their interview with us, this participant elaborated on why such warm-up activities were often necessary for ice-breaking in VR and how they were effective: 

\begin{quote}
    \textit{``The collective handclapping is kind of a bonding act. It borrows from gaming principles when other hosts and I design it, even though it's not an official game. People are not really clapping for a campaign in League [during that]. The focus is more on getting people to interact from scratch. It's like if someone yawns then everyone else yawns. It's mostly to show you are engaged with each other, and you are present together in a way where you don't have to talk, but you could. You wouldn't do much like this at a real-world meeting, but that's what makes VR special. You can do things to bond that are considered casual and maybe even goofy.''} [P13]
\end{quote}

Besides collective handclapping, avatar cloning was another popular warm-up activity we observed, unique to VRChat. Participants in the activity were invited to copy the public avatar used by their event host until they all looked the same. This temporary yet powerful sense of affinity helped people let down their guard and start talking to each other more easily. As P2 noted: 

\begin{quote}
    \textit{``The cloning activity was pretty fun. You can immediately switch to another avatar, and we all ended up looking exactly the same. I like it because it was like showing the people you're interacting with a kind of comradery. [This feeling is created at a] really low cost and [it's] very helpful for [fostering] more interactions. It's something special you can only do in VR.''} [P2]
\end{quote}

\textbf{In sum}, findings under the current section indicate that, among all three steps toward initiating opportunistic interactions, ice-breaking was arguably the most challenging one to navigate, especially when interactions occurred in social VR. The range of non-verbal cues, as well as the verisimilitude of those cues, are often insufficient for people to engage in small talk effectively and comfortably. Event hosts have experimented with various methods to facilitate ice-breaking among individuals. We observed many successful instances of their efforts, but also noticed an imbalance in the burden placed on hosts compared to attendees. 

\subsection{Acquiring Established Interaction Protocols in Social VR}\label{acquiring protocols}
Our participants reported referring to behavior protocols from real-world professional events to guide their interactions within the virtual environment. Under this general approach, many noted that, when translated into VR, these behavior protocols would inherently involve both social and technical dimensions.

Participants shared anecdotes witnessed from past events, where fully fluent social VR users could replicate their behaviors exactly as they would in physical reality. For example, those users could demonstrate natural body wiggling via their avatars. They sometimes even chose to leave virtual conference rooms by walking outside and then closing the virtual door, showcasing a sophisticated understanding of the hand controller mechanics as well as the virtual platform's specific features. Participants commented that such \textit{``super-users''} might always be adept at overcoming constraints in VR and recovering the verisimilitude of all meaningful cues. However, none of our participants identified themselves as such. Even for participants with extensive experience hosting professional events in VR, they were confident in their \textit{``VR fluency''} on the platforms they regularly used, but also highly aware that the technical protocol guiding people's action possibilities could vary significantly across platforms. 

Dedicated learning could be an intuitive solution for everyone to acquire the dos and don'ts at VR-mediated professional events. Our observations suggest that different social VR platforms all have offered instructional resources, ranging from text descriptions to video tutorials, to explain their technical features and the social functions they afford. Yet, somewhat predictably, few participants in our sample had ever consulted these resources, as they felt it would be \textit{``complex,''} \textit{``time-consuming,''} or \textit{``boring.''} Hosts often ended up being the on-the-ground resource, offering personal guidance to those in need of help. Several remarked sharply, \textit{``People may assume our job is to block bad behaviors. In reality, we do far more to facilitate good behaviors, and that requires far more work.''}

\begin{figure}[h]
    \includegraphics[width=1\columnwidth]{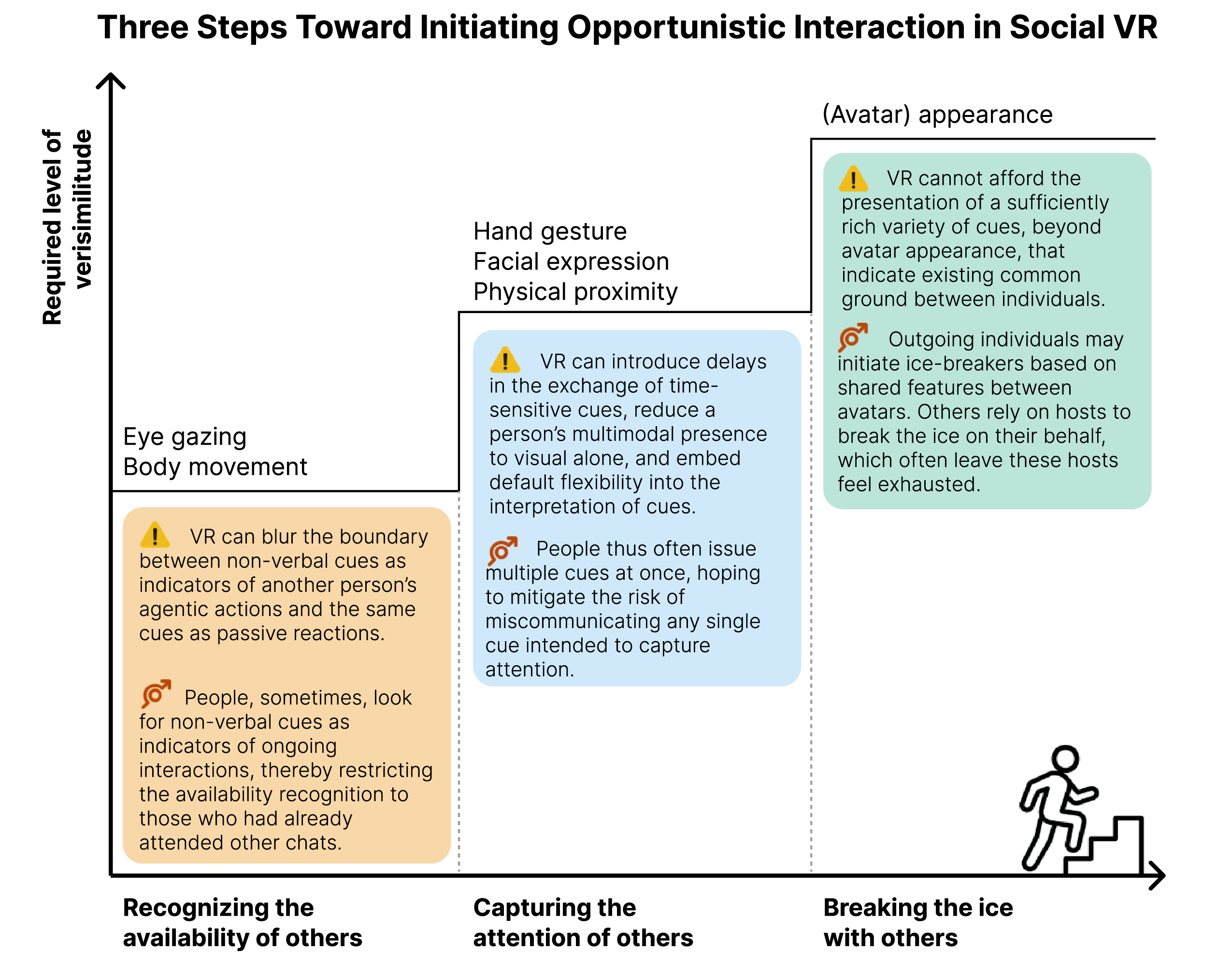}
    \caption{\edited{In face-to-face interactions, multiple non-verbal cues can be used simultaneously for a brief duration to initiate engagement. In contrast, initiating similar opportunistic interactions in professional social VR generally involves three steps, each requiring a progressively higher level of verisimilitude. Here, we summarize the challenges at each step, including the causes of difficulty in achieving verisimilitude and how people currently respond in practice.}}    
    \Description{In face-to-face interactions, multiple non-verbal cues can be used simultaneously for a brief duration to initiate engagement. In contrast, initiating similar opportunistic interactions in professional social VR generally involves three steps, each requiring a progressively higher level of verisimilitude. Here, we summarize the challenges at each step, including the causes of difficulty in achieving verisimilitude and how people currently respond in practice.}   
    \label{fig:summary}
\end{figure}

\section{Discussion}

To recap, our work illustrates that, at professional events in social VR, initiating opportunistic interactions is challenging and should not be regarded as a singular process as in real-world settings. Rather, \edited{it comprises three successive steps. Each step features the intuitive adoption of a selected set of non-verbal cues, and it requires a progressively higher degree of verisimilitude than the preceding one(s) to distill actionable meanings from these cues (see Figure~\ref{fig:summary} for a summary).} 
The technical infrastructure of today’s social VR platforms, unfortunately, often fails to support non-verbal exchanges with such subtleties. In this section, we elaborate on \edited{the empirical contributions as well as} the design implications of the current research. Our reflections center around three primary takeaways derived from our findings: \edited{the challenges in estimating verisimilitude for initiating opportunistic interactions (Section~\ref{escalating needs}), the unique role served by each stakeholder in initiating opportunistic interactions (Section~\ref{hosts, attendees, virtual agent}), and the preparation work that can be done in advance to ready people for initiating opportunistic interactions  (Section~\ref{training}).}

\subsection{Verisimilitude \edited{and Challenges in Estimating It}}\label{escalating needs}

\edited{Verisimilitude, the central notion connecting various segments of our data, underscores the fact that established social meanings attached to non-verbal cues may not fully transfer from face-to-face to virtual worlds. Multiple features of social VR can jointly influence how individuals interpret and act upon these cues for opportunistic interactions and, notably, they often do so in conflicting ways. On the one hand, the default visual design philosophy of having eyes, faces, and bodies in social VR mimic their real-world counterparts (mis)lead people to assign these non-verbal cues identical social meanings across worlds. On the other hand, exchanging non-verbal cues in social VR is, by nature, a mediated action. The VR platform usually has to replace complex social mechanisms with feasible technical ones to function (e.g., platforms involved in our study all leverage emojis to indicate the user’s facial expressions), and it swaps out preconditions that are necessary for a cue’s social meanings to remain intact (e.g., emoji selection takes time, which disrupts the chronemics). }

\edited{Most participants in our study lacked a sophisticated understanding of how social VR swapped out any preconditions for meaning interpretation at each moment toward opportunistic interaction. Nor could they always possess effective coping strategies to adapt to this shift. While some prior literature has also examined non-verbal cues in social VR-based interactions (e.g., ~\cite{maloney2020falling, maloney2020talking}), participants in those studies demonstrated creative ways of utilizing and even repurposing non-verbal cues in VR. They appeared far less concerned about VR’s disruptions to the verisimilitude of those cues than our participants did. We believe these disparities are rooted in differences between the context of prior work and ours: the former primarily explores social VR meetups for leisure and recreational purposes, whereas ours specifically focuses on professional networking. Participants in this latter context were reluctant for their behaviors in social VR to deviate significantly from the convention of real-world professional interactions. Precisely estimating the verisimilitude of non-verbal cues in social VR, when possible, helps people interpret others’ status in preparation for an appropriate initial interaction with them. Over the full course of a VR-based interaction following its initiation, we suspect that people will continue to leverage their understanding of verisimilitude to calibrate other important aspects of communication, such as which impression management strategies to employ and how much reciprocal exchange is needed to reduce uncertainty.}

\edited{So, is there a low-effort solution to make the estimation of verisimilitude an easy task for everyone? The answer, based on our current findings, leans toward \textit{no}. VR environment design is often a compromise among fidelity against real-world settings, technical feasibility, and the choices and preferences imposed by platform designers. This context makes it inherently difficult for non-verbal cues to carry equivalent meanings across social VR and the real world. Therefore, before exploring ways to enhance a person’s estimation of verisimilitude at any given moment, it may worth stepping back to ask whether non-verbal cues displayed during social interactions are, in fact, the ideal means for people to recognize one another's status in VR.}

\begin{figure}[h]
    \includegraphics[width=0.85\columnwidth]{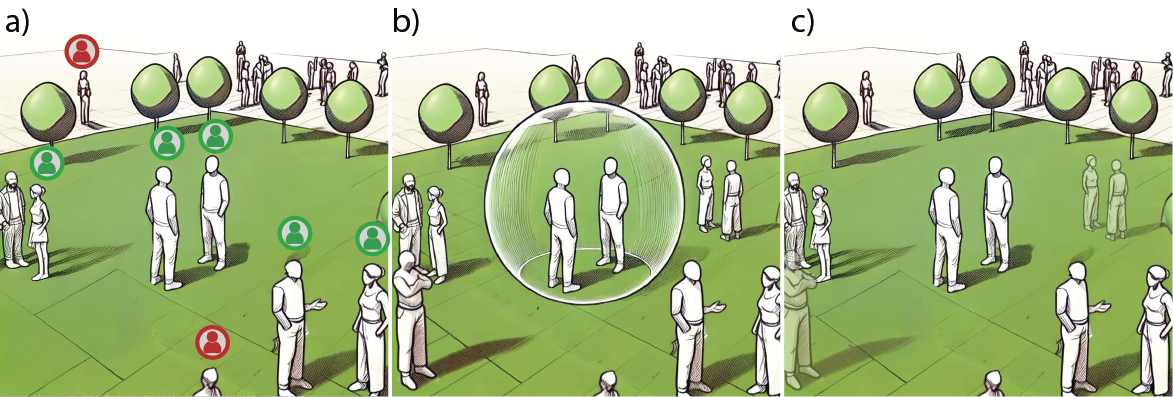}
    \caption{Three possible ways to indicate a VR user's availability or willingness to engage in conversation: a) status indicators as floating toggle icons above the avatar; b) bubbles as visual cues that clearly indicate chat availability; c) fading out avatars that are currently unavailable for social interaction. Illustrations generated by DALL-E 3 based on a user-provided prompt. }    
    \Description{a) Status indicators as floating toggle icons. b) Bubbles that clearly indicate chat availability. c) Fading out of avatars that are not available. Illustrations generated by DALL-E 3 based on a user-provided prompt.}   
    \label{fig:availability}
\end{figure}

\edited{The question above implies that there can be ways to walk around the complexity of estimating verisimilitude while still arriving at a satisfying understanding of another person’s status.} One straightforward means, for example, to is to provide social VR users with a simple and clear indicator of others’ availability or willingness to engage in conversation. Such an indicator could take various forms: it might resemble familiar toggle icons from instant messaging tools~\cite{maloney2020falling, maloney2020talking}, or more immersive options, such as bubbles surrounding avatars to indicate willingness to chat, or a filter that fades out avatars unavailable for interaction (Figure~\ref{fig:availability}). When a person is open to opportunistic chatting with others, they can opt in to display this status indicator publicly.

\edited{Here, we see a critical trade-off when comparing non-verbal cues and artificial status indicators as distinct mechanisms for interpreting another person’s status in VR. On one side of this trade-off is the design vision of enabling social VR to support natural human interactions as if they were taking place in the real world. On the other side is the practical need to reduce users’ effort in navigating these interactions. Findings from our current research suggests that it can take numerous trials and errors before a person develops an effective mental model about the verisimilitude of non-verbal cues within each social VR environment. For platform operators whose top priority is to facilitate successful interactions at VR-based professional events, incorporating artificial status indicators into their system design appears to be a reasonable choice, even though it will limit the natural unfolding of social meanings.}

\subsection{\edited{Social Responsibilities} Delegated to the Attendee, the Host, or \edited{the System}}
\label{hosts, attendees, virtual agent}
\edited{Estimating the verisimilitude of non-verbal cues in social VR is a complex task, and failures in this estimation can carry significant social costs in the context of professional events. Thus, rather than assuming that individuals as event attendees must solve this verisimilitude puzzle on their own, an alternative perspective is to consider who else can be entrusted with similar responsibilities and how they may coordinate with attendees to distribute the effort effectively.} 

\edited{In this regard, much of our research highlights the need to better understand and support the actions taken by event hosts. When attendees at professional events were uncertain about how to initiate opportunistic interactions, hosts often act as social lubricants — for example, by inviting multiple attendees to participate in the same warm-up activity — lowering the barrier people would otherwise encounter to extended interaction. Our findings offer a valuable complement to prior work on social VR, which has primarily examined it as an emerging medium for recreational purposes. In these prior settings, behavioral norms are often loosely communicated or, at least, features a high degree of ambiguity. Event hosts are typically responsible for identifying and correcting toxic or disruptive behaviors performed by specific individuals (e.g., ~\cite{blackwell2019harassment, schulenberg2023we}). The current research, in contrast, sheds light on a broader and more proactive role for hosts in social VR. Instead of focusing on policing a problematic few, hosts in the context of professional events actively worked to foster inclusive and smooth interactions among all attendees. The effort required to fulfill these responsibilities, as reported by our participants, is substantial. }

\edited{Nevertheless, the current research reveals a ``responsibility-ability'' gap experienced by hosts. Our data do not show systematic differences between hosts and attendees in their abilities or confidence in estimating the verisimilitude of non-verbal cues, especially when considering participants’ self-reports during the availability recognition and attention-capturing steps. In fact, multiple participants had experiences being hosts and attendees at different events, respectively. While they demonstrated general awareness that verisimilitude in social VR can be puzzling, we cannot conclude that they had developed more sophisticated coping strategies than others for recovering from disrupted verisimilitude. The advantage of being a host is more evident at the ice-breaking step. In that context, the host role implies a general expectation, making it socially appropriate for these individuals to initiate small talk or propose warm-up activities at any time. To some extent, this expectation exempts hosts from the social pressure to interpret non-verbal cues with high precision. Their unique social status grants them permission to experiment with various methods to facilitate ice-breaking.} 

\begin{figure}[h!]
    \includegraphics[width=0.6\columnwidth]{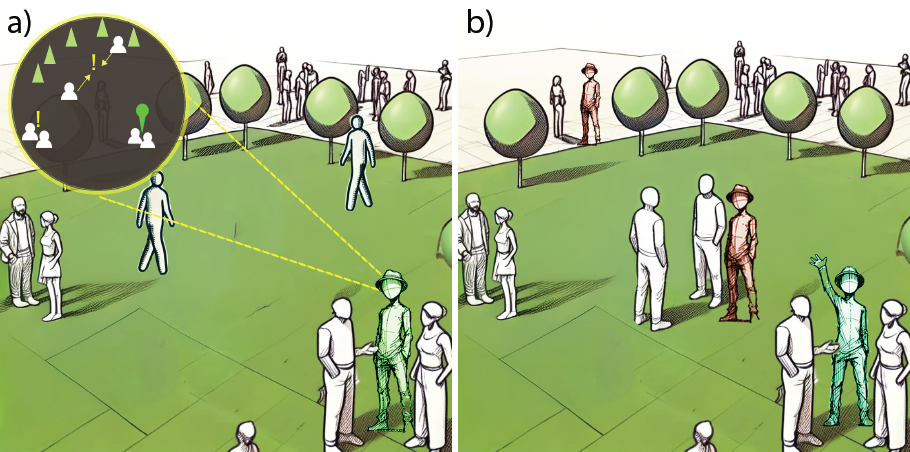}
    \caption{Hosts and non-human embodied agents can help facilitate opportunistic conversations. a) VR platforms may notify hosts (shown in green) of potential hotspots where such conversations are likely to emerge. b) Embodied conversational agents (shown in red) can recommend icebreakers based on context, event topics, or user profiles. Illustrations generated by DALL-E 3 based on a user-provided prompt.}
    \Description{Hosts and non-human embodied agents can help facilitate opportunistic conversations. a) VR platforms may notify hosts of potential hotspots where such conversations are likely to emerge. b) Embodied conversational agents (shown in red) can recommend icebreakers based on context, event topics, or user profiles. Illustrations generated by DALL-E 3 based on a user-provided prompt.}    
    \label{fig:map_agent}
\end{figure}

\edited{The above reflections point to possible ways through which the effort for initiating opportunistic interactions could be partially delegated to parties other than the attendees themselves. The most obvious direction is to incorporate design features that consolidate the host’s existing practices of breaking the ice for the attendees’ benefit. Given that professional events often place multiple demands on hosts’ attention, facilitating ice-breaking can only happen when these individuals are available at the right moment, have identified a potential opportunistic interaction among attendees, and are nearby. All these constraints introduce considerable randomness into the timing and format of a host’s performance.} One feasible support from the system end is to automatically identify situations where ice-breaking assistance is needed at an ongoing event. For example, the VR platform could notify hosts of potential locations, or hot spots, where opportunistic conversations might occur between participants, allowing hosts to teleport precisely and provide facilitation (Figure~\ref{fig:map_agent}a). To implement this functionality, the system can leverage real-time location data of all attendees. The likelihood of an opportunistic conversation can be inferred by computing proximity between attendees and analyzing each person’s movement trajectories, (e.g., ~\cite{cranshaw2010bridging, zheng2010geolife}).

Along the same direction but taking another step further, we envision that the facilitation effort, currently undertaken by human hosts, may in the future be delegated to embodied virtual agents who fulfill similar social responsibilities but are non-human (Figure~\ref{fig:map_agent}b). Such agents could appear as host avatars beside attendees, encouraging ice-breaking through LLM-based conversational abilities. \edited{Much research on social matching tools, initially proposed by Terveen and McDonald~\cite{terveen2005social}, has leveraged  user profiles and interests~\cite{park2024who2chat}, ad-hoc encounters~\cite{song2021online}, contextual data~\cite{mayer2016supporting}, or interaction history~\cite{im2020synthesized}, for recommending interaction opportunities of the user's interest. We believe the similar information can be used in professional VR events, where attendees may voluntarily provide details such as their professional background, event objectives, and desired networking contacts during the event.} This data can then be input into a matching algorithm, which can then recommend conversation topics tailored to personal preferences. The virtual agents, acting as host replicas, would then prioritize the recommended topics while facilitate ice-breaking conversation. Notably, taking this approach requires researchers and system builders to proactively think through the potential social implications of AI-facilitated interactions (e.g., ~\cite{guimaraes2020impact, wan2024building}). To maintain the authenticity of human interactions, we believe the involvement of virtual agents should not undermine agency and accountability of human hosts. Attendees must, at the very least, be clearly informed when interacting with non-human agents. \edited{More empirical work will be needed to enrich our understanding of what social consequences may arise in such a mixed-initiative context and how to communicate these consequences for supporting human actions.}

\subsection{In Preparation for Successful Opportunistic Interactions at Social VR Events}\label{training}
\edited{Up to this point, we have discussed using artificial status indicators and delegating partial effort to hosts (and host-like virtual agents) as two possible ways to work around existing challenges in estimating verisimilitude for opportunistic interactions. We now turn to a more difficult yet crucial question: are there direct ways to enhance a general social VR user’s ability to interpret and act upon this verisimilitude effectively? }

\edited{Our research findings acknowledge that participants at social VR events have been accumulating their knowledge to formulate some mental models about what boosts or disrupts verisimilitude in the virtual environment. The problem, though, is that their knowledge is largely experiential. Each VR platform follows its own rules in regulating how gazing, hand clapping, and avatar-based presentation work. Each individual also holds their incomplete assumptions about how much real-world social meaning could be, or should be, assigned to non-verbal exchanges in VR. Such variations make a person’s experiential knowledge about verisimilitude highly vulnerable to context shifts. As people move from one event to another, they must test the transformability of their established mental models through yet another taxing round of trial and error.}

Forcing a greater level of consistency in the system-level operation of non-verbal cues could be one way to reduce variations. Specifically, we imagine that a unified design policy could be implemented to govern the exchange and interpretation of non-verbal cues across different VR platforms and to align relevant behavioral protocols among all users. This design policy may function similar to how emoji usage has become increasingly standardized through Unicode across Android, iOS, and other web applications. 

\edited{We suspect that the lack of a unified policy in current social VR platform design is not due to feasibility concerns; rather, it is a deliberate choice shaped by dominant use cases in the past. As discussed earlier, VR has long been considered a medium for exploring creative forms of casual encounters (e.g.,~\cite{freeman2021hugging}) and for gaming and social grooming among young generations (e.g.,~\cite{maloney2020virtual}). Much of the engaging and joyful user experience in these contexts is arguable tied to the unconventional and un-unified aspects of VR environment design. The recent adoption of social VR for professional events during and post the COVID pandemic pushes researchers and system builders to revisit design guideline for this medium, particularly in striking the right balance between innovation and real-world resemblance for effective communication of non-verbal cues. 
Independent of the idea of making the VR environment more structured for users’ mental model formation, another approach is to facilitate people in distilling context-sensitive knowledge from the trial and error they will inevitably go through. Some of our research findings have shed light on this second approach.} Specifically, participants in the current study recognized that system tutorials and user manuals, both available via the platform they used, were valuable and legitimate resources to get themselves familiar with specific system features and the social functions being afforded. However, they preferred not to leverage these standalone resources, instead choosing to accumulate experiential knowledge through real interactions with others. 

\begin{figure}[h!]
    \includegraphics[width=0.6\columnwidth]{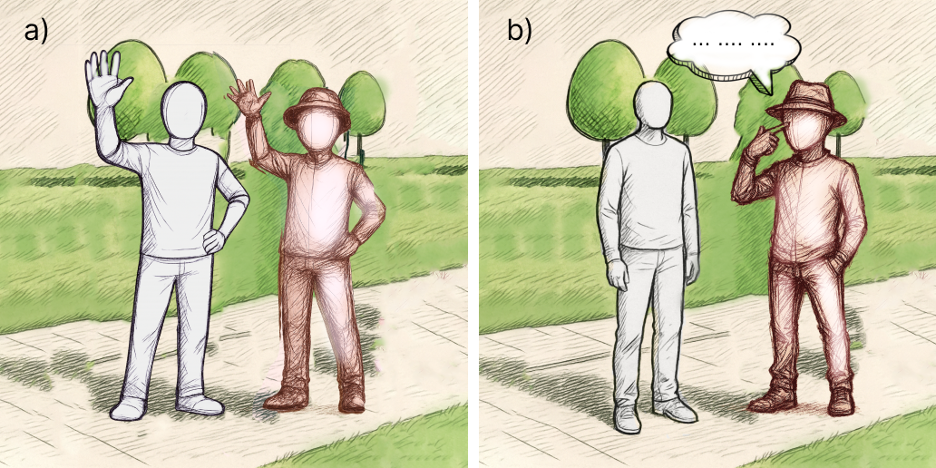}
    \caption{Future VR platforms may consider embedding system tutorials directly within social events hosted on the platform. a) Users can familiarize themselves with ways to issue non-verbal cues through situated trial-and-error interactions with a virtual agent (shown in red). b) Virtual agents can also provide dynamic feedback to guide users in making reliable and context-sensitive interpretations of a non-verbal cue’s social meanings. Illustrations generated by DALL-E 3 based on a user-provided prompt.}
    \Description{Future VR platforms may consider embedding system tutorials directly within social events hosted on the platform. a) Users can familiarize themselves with ways to issue non-verbal cues through situated trial-and-error interactions with a virtual agent (shown in red). b) Virtual agents can also provide dynamic feedback to guide users in making reliable and context-sensitive interpretations of a non-verbal cue’s social meanings. Illustrations generated by DALL-E 3 based on a user-provided prompt.}    
    \label{fig:training}
\end{figure}

The observations above suggest that future VR platforms should consider embedding system tutorials directly within social events hosted via the platform, following practices already proven successful in the gaming field (e.g., ~\cite{adams2012game, white2014learn}]). \edited{Through such situated trial-and-error experience, individuals can effectively form episodic memories about how to issue various non-verbal cue in a given VR environment, how this non-verbal exchange differs from its real-world counterparts, and whether any alternative mechanisms might help achieve the similar social purposes ({Figure~\ref{fig:training}}). Essential knowledge for the estimation of verisimilitude is, thereby, accumulated in a contextualized and system-governed manner, offering the potential to prepare diverse users for successful opportunistic interactions at social VR events.}

\section{\edited{Reflections on Research Ethics}}

\edited{In addition to discussing our primary findings, we here dedicate a separate section to reflecting on ethics-relevant decisions made before and over the course of this research. Consenting to research data collection at social events involves complex considerations — for instance, who should provide direct permission for their self-reports and observational data to be documented, what information should be disclosed to others interacting with those providing consent, and to what extent, and in what ways, researchers should make their own presence salient at such events. 
We begin our reflections by discussing the rationale guiding our decisions during the research preparation stage. We then summarize insights shared by previous scholars who have explicitly proposed criteria for qualitative data collection in both face-to-face and virtual settings (see ~\cite{chang2025ethical} for a more comprehensive review). By introducing our own practices and situating them within broader ethical discussions, our aim is not to claim that we have done everything “correctly.” Rather, we hope readers of this paper, like ourselves, can recognize “grey spaces” that still persist in defining and implementing ethical data collection in social VR. For those with similar interests in studying interactions in VR, we offer these reflections as a resource to support the development of their data collection protocols, which can hopefully balance various ethical considerations.}

\edited{\minor{The protocols described in Section~\ref{protocol} "Observation and Interview Protocol" } were designed to maximize bystanders’ awareness of the researcher’s presence and to prevent their behaviors from being observed without meaningful transparency, despite the fact that all events being studied in the current work were already public in nature. Our approach echoes recent ethical discussions on social VR research (e.g., ~\cite{maloney2021social, cockerton2024conducting}), which emphasize the importance of making data collection processes transparent to VR users, the need to align the researcher’s practices with the community norms of participants, and the responsibility to protect the anonymity of non-consenting users. It also consulted more established protocols for real-world qualitative research in sociology and relevant fields. In particular, the guidelines set by the American Sociological Association’s Scope of Informed Consent~\cite{ASA1997}, in which they explain in section 12.01(c) that researcher “may conduct research in public places or use publicly-available information about individuals (e.g., naturalistic observations in public places, analysis of public records, or archival research) without obtaining consent,” and should “disguise the identity of research participants…or other recipients of their service” if confidential information is collected.}

\edited{While the stated protocols appeared to work well for our research purposes and did not violate any strict ethical guidelines, we do not and cannot claim that they represent the best practices for human data collection in social VR. One important insight we gained from reviewing broader literature on research ethics is that data collection at public events—whether in face-to-face or virtual settings—can be jointly governed by multiple, sometimes competing, ethical criteria (e.g.,~\cite{spicker2011ethical, power1989participant, waern2016ethics, marzano2021covert, podschuweit2021ethical, li2008ethical}). Some HCI scholars, for example, have explicitly called for careful trade-offs among the ecological aspect (e.g., will my consent process disrupt the phenomenon being studied?), the normative aspect (e.g., does my consent process align with what people would reasonably expect?), and the practical aspect (e.g., is my consent process feasible to implement?), among others (e.g., ~\cite{williamson2016deep}). Researchers in sociology and related social sciences sometimes adopt a more moderate stance, arguing that in research which is “less interventionist” and entails “less emphasis on risk management than that found in (clinical) psychology and medicine,” the notion of an appropriate consent process may reasonably allow for a range of approaches. }

\edited{As we look back on our research practices again through the lens discussed above, it becomes clear that some of our decisions could have been made differently. These hypothetical alternatives may better satisfy certain ethical criteria while raising concerns against others in their own ways. For instance, instead of implementing different consent processes for registered participants and those interacting with them, researchers might conduct a comprehensive consent process with everyone attending the event to place them on equal footing. It could also be valuable to co-develop user-controlled mechanisms that prevent the researcher from eavesdropping on conversations. We did not explore this possibility in the current study but relied on a trust-based approach, expecting the researcher to follow a “code of honor” and adhere to their commitments made to participants.}

\edited{So why didn’t we adopt these alternatives in the first place? The honest answer is that we were uncertain about the level of effort required to implement them and whether that effort would be worthwhile for the participants and for the research team. In a public online event that is technically open to anyone and allows people to log in and out freely, we were concerned that an equal consenting approach would place heavy coordination burdens on the host. As for the issue of eavesdropping, delegating control to participants might have led to frequent interruptions to their experience at the event, diverting attention from their networking with others.}

\edited{We imagine that, as human-subject research becomes increasingly common in social VR, future systems will offer user profile settings specifically designed to support research activities. For instance, when an avatar with a “researcher profile” enters some event hosted in social VR, the system could trigger a series of automated actions to obtain informed consent from all attendees. Reporting detailed data collection protocols, along with associated ethical measures, in research publications is a crucial step toward the effective design of such system features. This practice can also promote greater exchange among research groups and contribute to building community-wide consensus around ethical standards and guidelines.}

\section{Limitations}
Our research findings should be interpreted with certain limitations. By employing a qualitative approach that included in-situ observations and in-depth interviews, we gained nuanced insights into how participants initiate opportunistic interactions at professional social VR events. 
However, this approach limited the number of participants we could include in the study. 
While we aimed to examine professional events across diverse platforms and involved participants in various roles, our final sample may not represent the full range of situations relevant to social interactions in VR. 
\edited{For instance, all participants in our study located in the United States, although the observed events were hosted in virtual platform and could, technically, involve individuals from diverse national cultures. We collected certain aspects of their background information, as detailed in Table~\ref{table:1}, but it did not cover an exhaustive list of details, such as the person’s job title or position in the organizational hierarchy, number of years in their professional field, and many others. These constraints made the exploration of possible associations between the person’s social VR behaviors and their dispositional factors unlikely. Furthermore, }
the researcher's presence may have subtly influenced participant behavior, although this involvement was necessary to capture the level of detail we sought while allowing participants control over their disclosures.
\edited{We look forward to future studies that can cross-validate and expand our findings by examining VR-based opportunistic interactions via other methods, including quantitative modeling for social VR user behavior prediction, in-situ tracking to detail multiple users’ interaction trajectory over time, and longitudinal research to understand whether interactions initiated in social VR have any carry-over effects on real-world communication among the same group. In this regard, our current work has contributed essential concepts and their contextualized associations for future research to take up.  }

\section{Conclusion}
Opportunistic interaction serves as the starting point for social bonding and information sharing in professional settings. Managing such interactions effectively is not always easy in physical environments and could be even more challenging on social VR platforms. In this paper, we presented a qualitative study aimed at understanding and enhancing opportunistic interactions at VR-based professional events. Our sample consisted of 16 individuals with ongoing experience in this space of interest. We conducted comprehensive observations at one or more events attended by each participant, taking detailed notes on their interactions at the events and conducting interviews to gather their self-reflections. Through inductive analysis of this data, we identified three distinct steps leading to successful opportunistic interactions at professional events in social VR: recognizing availability, capturing attention, and ice-breaking. The assessment of verisimilitude emerged as central to participants' actions, experiences, and decision-making across all three steps. At each step, we uncovered unique issues influencing how participants interpreted and acted upon the verisimilitude of non-verbal cues. Our work sheds light on the future design of social VR platforms for professional purposes, offering suggestions for revising human-VR interface design, enhancing host-attendee coordination, and creating situated educational opportunities for individuals from diverse backgrounds.

\begin{acks}
We are grateful to our participants for their time and insights, which made this work possible. We thank Yongle Zhang, Yimin Xiao, Cartor Hancock, and Yuhang Zhou, as well as the anonymous reviewers, for their valuable feedback on various versions of our paper.  We also thank Hao-Chuan Wang, John Tang,  and Jessica Vitak for earlier discussions that informed this work.
\end{acks}

\bibliographystyle{ACM-Reference-Format}
\bibliography{sample-base}

\received{October 2024}
\received[revised]{April 2025}
\received[accepted]{August 2025}
\end{document}